\documentclass[pra, twocolumn, floatfix]{revtex4}
\usepackage{graphicx}
\usepackage{amsmath, amsfonts, amssymb, bm}
\usepackage{braket}
\usepackage{color}
\usepackage[utf8]{inputenc}
\usepackage{subfigure}

\begin{document}

\title{ Probing the helium dimer by relativistic highly-charged projectiles 
} 
\author{ B. Najjari $^1$, Z. Wang $^{1,2,3}$,   
and A. B. Voitkiv $^2$ }
\affiliation{ 
$^1$ Institute of Modern Physics, Chinese Academy of Sciences, Lanzhou 730000, China \\  
$^2$ Institute for Theoretical Physics I, Heinrich-Heine-Universit\"at D\"usseldorf, Universit\"atsstrasse 1, 40225 D\"usseldorf, Germany \\ 
$^3$ University of Chinese Academy of Sciences, Beijing 100049, China  } 
\date{\today}

\begin{abstract} 
We study the fragmentation of He$_2$ dimers into He$^+$ ions by 
relativistic highly-charged projectiles. 
It is demonstrated that the interaction between  
an ultrafast projectile with an extremely extended object -- the helium dimer -- possesses interesting features which are absent in collisions with "normal"  molecules. 
It is also shown that such projectiles, 
due to their enormous interaction range,   
can accurately probe the ground state of the dimer 
and even be used  
for a precise determination of its binding energy.    
\end{abstract}

\maketitle

1. The helium dimer, He$_2 $, is a fascinating quantum system bound by van der Waals forces. 
The interaction between two ground-state helium atoms 
is so weak that it supports just one bound molecular state 
with a tiny binding energy 
($ \simeq 10^{-7}$ eV, \cite{e-l})  
and an enormous size: its average bond length is  
$ \approx 52$ \AA \, \cite{e-l} while the dimer extends to 
the distances of more than $ 200 $ \AA \, representing 
the largest known (ground-state) diatomic molecule. 

Because of such extreme dimensions,   
the Casimir-Polder retardation effect \cite{Cas-Pol} 
(having a relativistic origin) noticeably influences 
the interatomic interaction in the dimer ground 
state \cite{retard}.  
The outer classical turning point in 
this state is about $14$ \AA \, \cite{14A} 
that is almost four times smaller than 
its average size indicating that the dimer is a quantum halo system which spends most of the time in the classically forbidden region. 

Even though the possible existence of this dimer was 
theoretically discussed since the 1920th, 
it was experimentally observed only in 1993-1994 
\cite{He2-exp-1}-\cite{He2-exp-2}.    
   
\vspace{0.05cm}

2. When atomic particles (atoms, molecules) 
interact with each other or 
with external electromagnetic fields, they can be excited, 
ionized or even disintegrated. 
Ionization and fragmentation processes are of 
especial interest because 
they not only unveil valuable information about the initial system but also 
trigger various transformations in chemical and biological envirounments. 

\vspace{0.05cm}

The fragmentation of the He$_2$ dimer 
into He$^+$ ions induced by absorption 
of a single photon  
or by a collision 
with a charged projectile 
was explored in \cite{He2_photon}-\cite{He2_photon_icd} 
and \cite{He2_alpha-particle}-\cite{He2_S14+}, respectively. 
In these papers the focus was on the fragmentation events   
with kinetic energies 
of the detected He$^+$ ions  
$\gtrsim 1$ eV  
which corresponds to the start of the Coulomb explosion 
of the He$^+$--He$^+$ system at  
internuclear distances not  
exceeding $ \simeq 14$ \AA.  

The authors of \cite{dimer-binding-exp}
investigated the fragmentation of He$_2$ into He$^+$ ions  
either by absorption of high-frequency photons or by a strong low-frequency laseer field.   
In both cases the dimer atoms 
were singly ionized within 
a time interval shorter than the time scale 
of the nuclear motion in the He-He$^+$ ion, 
which enabled one to sample 
the dimer wave function 
and measure its binding energy.  

\vspace{0.05cm} 

3. Here we report on our study of    
the fragmentation of the helium dimer into 
singly charged ions by 
collisions with relativistic highly-charged projectiles.      
It will be demonstrated that the interaction of 
an ultrafast projectile with an extreme extended object -- the helium dimer -- possesses interesting (and exceptional) features which are absent 
in collisions with "normal" molecules (or atoms). 

It will also be shown that such projectiles, 
due to their ultralong interaction range,   
can directly probe the structure of the dimer in 
the halo region $\sim 14 - 250 $ \AA \, 
(where it spends about $ 80 \% $ 
of the time \cite{14A})  
and can be even used for an accurate determination of 
the binding energy of the dimer. 

Atomic units ($\hbar = \vert e \vert = m_e = 1$) 
are used throughout unless otherwise is stated.  

\vspace{0.1cm}

4. Let He$_2$ dimers in the ground state collide with  
projectiles having a charge $Z_p$ and moving 
with a velocity ${\bm v}$  
approaching the speed of light $c \approx 137$ a.u.  
In such collisions the parameter $\eta = Z_p/v$ always remains well below $1$ indicating that the unitarity condition does not "couple" different reactions  which may, therefore, be considered separately. Also,  
the inclusion of any extra interaction step 
(beoynd a necessary minimum) sharply 
reduces the production cross section. 
Besides, at $ v \sim c$ electron capture processes are  completely negligible 
compared to ionization \cite{el-cpt}. 
Under such circumstances  
the breakup of the He$_2$ dimer into He$^+$ ions 
caused by these collisions, 
\begin{eqnarray}
Z_p + \text{He}_2 \to Z_p + \text{He}^+ + \text{He}^+ + e^- + e^- ,   
\label{fm1-1}
\end{eqnarray} 
is strongly dominated by the following 
fragmentation mechanisms.  

i. First, the projectile "simultaneously" 
interacts with both atomic sites of the dimer. As a result, each helium atom emits an electron becoming 
a singly charged ion \cite{Sulf}. 
Since in this fragmentation mechanism the projectile directly 
forms the transient He$^+$ - He$^+$ molecular 
ion, while the interactions between the constituents of 
He$_2$ play here no noticeable role, we shall call it 
{\it the direct fragmentation} (DF).  
Due to the repulsion,   
the He$^+$ - He$^+$ system is unstable 
undergoing a Coulomb explosion.  

The reflection approximation 
relates the kinetic energy $ E_{ker} $ 
released in the explosion to the internuclear distance 
$R_{ce}$ at which it started: $ E_{ker} = 1/R_{ce} $ 
\cite{RA}. 
On the time scale of the nuclear motion 
in the dimer the DF mechanism leads to the 
sudden removal of two electrons  
and $R_{ce}$ coincides with the instantaneous size 
$R$ of the dimer when the collision occurred.   

ii. Second, the projectile interacts with just one atom 
of the dimer, 
the atom is singly ionized and  
the emitted electron moves towards the 
other atom knocking out one of its electrons. 
This mechanism is a combination of single ionization 
of a helium atom by a high-energy projectile 
and the so called e-2e process on helium 
(single ionization by electron impact)  
and will be abbreviated as the SI--e-2e \cite{f1}.    

In the SI--e-2e     
the emitted electron moves much faster than 
the helium nuclei. Consequently, in this mechanism   
(like in the DF) 
the energy $E_{ker}$ directly reflects 
the dimer size at the instant of the collision 
with the projectile. 

iii. Third, the projectile also interacts with just 
one helium atom but now this results in 
its ionization-excitation. The residual helium ion then 
de-excites by transferring the energy 
to the other atom that leads to its ionization.  

iv. The last -- fourth -- mechanism also involves 
a collision of the projectile with just one atom 
which, in this case, 
leads to its double ionization. 
Then the He$^{2+}$ ion radiatively captures 
one electron from the neutral atom.    

Results on the break-up of He$_2$ dimers by 
photo absorption \cite{He2_photon_icd} and 
$0.15$ MeV/u alpha particles \cite{He2_alpha-particle} 
show that in the last two fragmentation mechanisms    
a very significant contraction of the 
He-(He$^+$)$^*$ and He-He$^{2+}$ dimers, 
which are directly produced in the collision,   
is necessary for the formation of the He$^+$-He$^+$ system. 
This is especially true for the mechanism, 
which involves the electron transfer occurring at small 
internuclear distances $R$ ($ \lesssim 2$ \AA, 
$E_{ker} \gtrsim 5$-$7$ eV \cite{He2_alpha-particle}) 
while the range of the two-center energy exchange 
is mainly restricted to $R \leq 14$ \AA, 
$E_{ker} \gtrsim 1$-$2$ eV \cite{He2_photon_icd}. 
In relativistic collisions, where the projectile field can be represented by "equivalent photons" \cite{WW},  
these mechanisms will possess same features as just mentioned above.  
  
By comparing the fragmentation mechanisms 
we thus see that only in 
the DF and SI--e-2e   
the energy $E_{ker}$ is directly   
related to the size $R$ of the dimer at 
the collision instant. Therefore, 
unlike in the other two, 
in the DF and the SI--e-2e mechanisms the ground state of the dimer is directly probed. Moreover, they can be separated from the other ("interfering") two by focusing on fragmentation events with small $E_{ker}$ ($E_{ker} < 1$ eV) and in what follows we shall consider them only.

5. The DF mechanism.   
Its detailed theoretical description will be given elsewhere. Here we just mention that 
it predicts that the DF cross sections 
depend on the projectile charge ($ \sim Z_p^4$), 
the impact energy (per nucleon) 
and the transverse size, 
$R_\perp = \sqrt{R^2 - ({\bm R} \cdot {\bm v})^2/v^2}$, 
of the dimer, where ${\bm R}$ is the dimer internuclear vector.  

\begin{figure}[h!]
\vspace{0.2cm}
\centering
\includegraphics[width=9.cm]{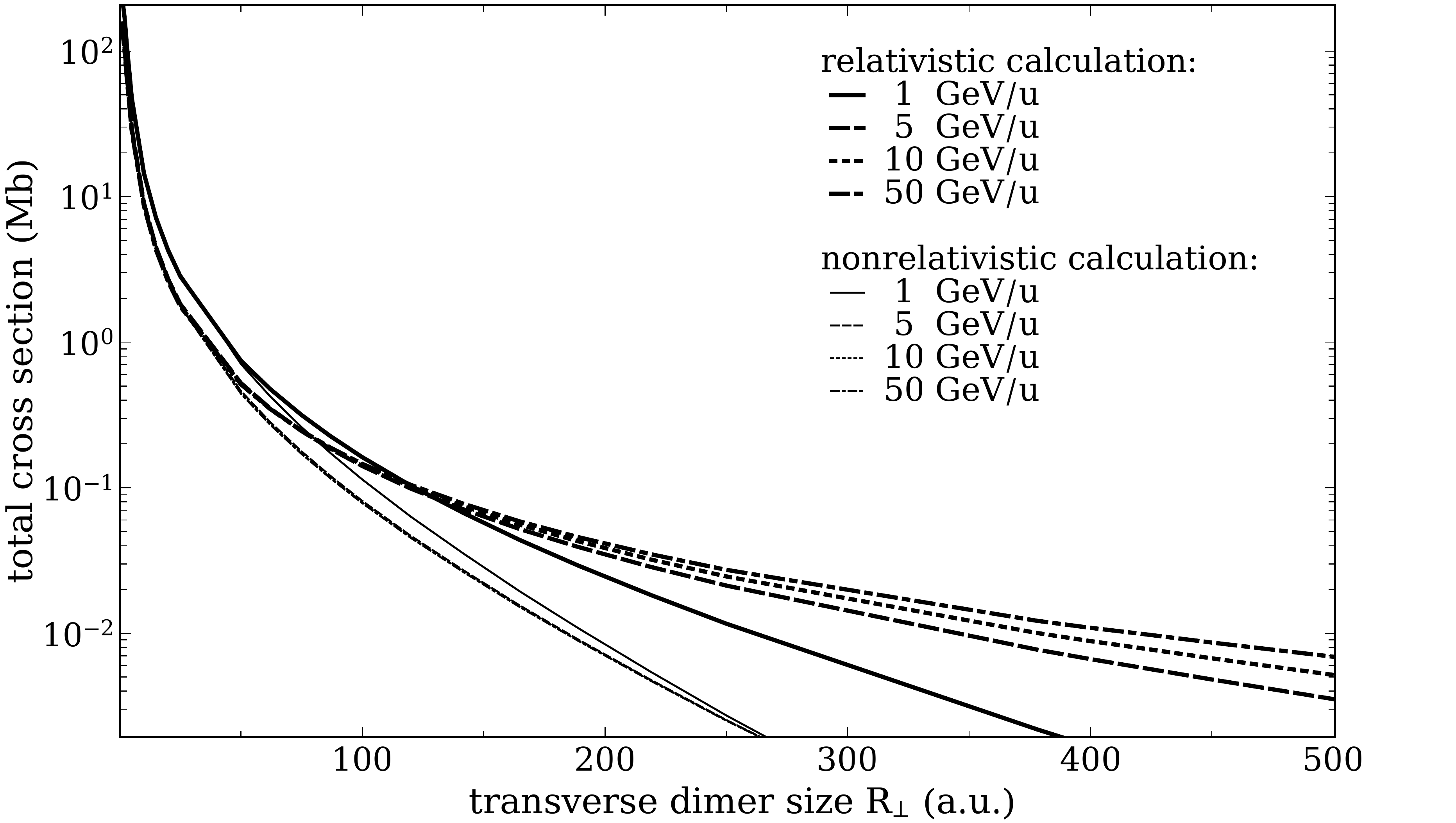} 
\vspace{-0.55cm}
\caption{ Cross section for producing  
two He$^+$ ions by U$^{92+}$ projectiles 
via the DF as a function of $R_\perp$. (Note that at 
$\gtrsim 5$ GeV/u all nonrelativistic results practically coincide.)}
\label{figure1}
\end{figure}

\begin{figure}[h!]
\vspace{-0.15cm}
\centering
\includegraphics[width=9.cm]{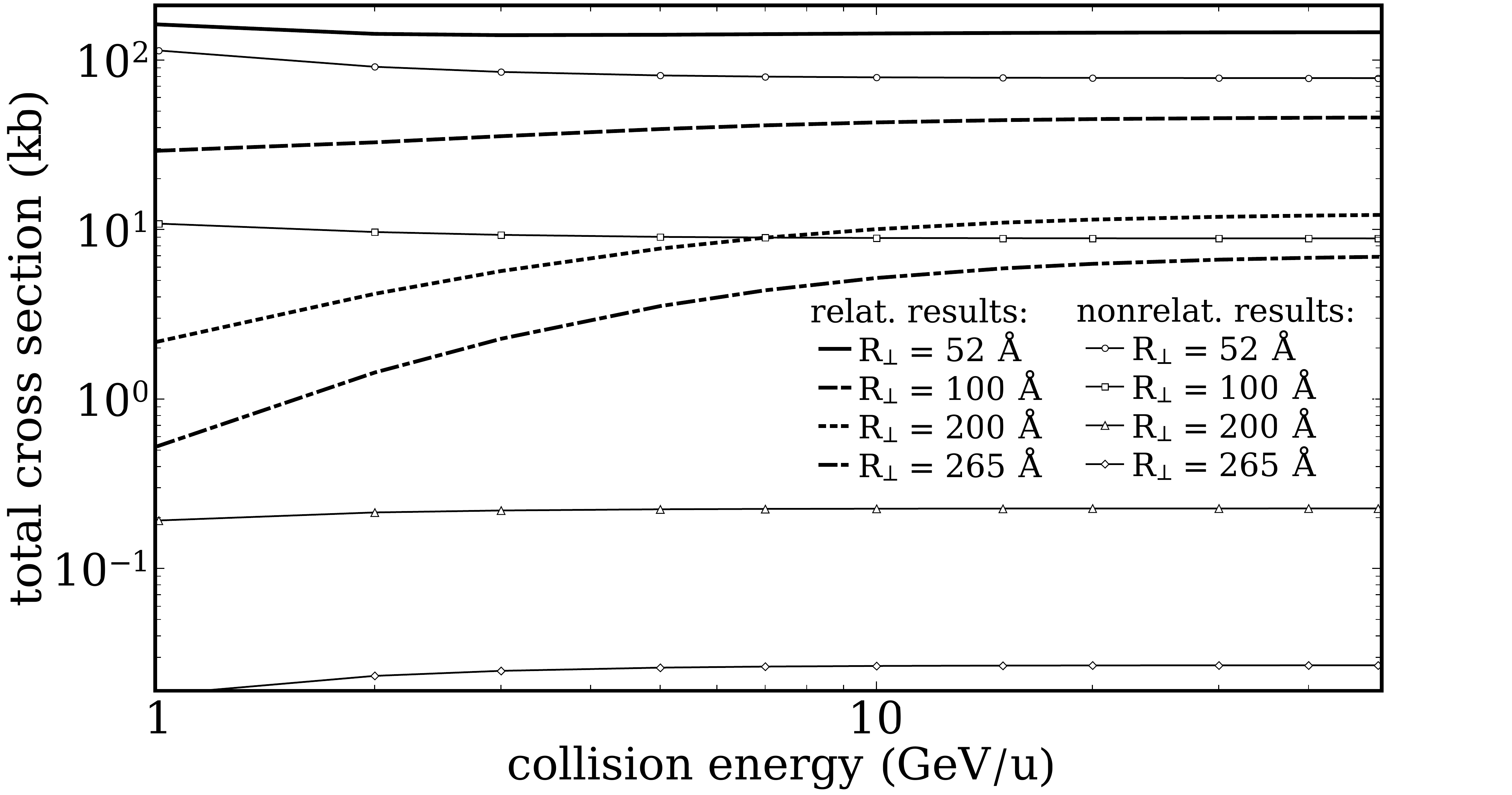} 
\vspace{-0.55cm} 
\caption{ Same as in figure \ref{figure1} but 
as a function of the impact energy.}
\label{figure2}
\end{figure}

In figures \ref{figure1}--\ref{figure2} 
we show the cross section 
for the production of two singly charged helium ions  
by U$^{92+}$ projectiles \cite{Z-v}. In figure \ref{figure1} 
the cross section is given as a function of 
the transverse size $R_\perp$  
at different impact energies 
whereas in figure \ref{figure2} 
it is plotted as a function  
of the impact energy for 
different values of $R_\perp$. 
Some main conclusions can be drawn from these figures. 

First, the cross section is quite sensitive to the transverse size of the dimer varying by orders of magnitude (with the sensitivity becoming less strong when the impact energy increases).    
However, even for the less favourable collision geometry, when $ {\bm R} \perp {\bm v}$ and the 
instantaneous size $R$ of the dimer is close to its maximal  detectable value ($\simeq 200$ - $250$ \AA,  \cite{dimer-binding-exp}),   
the cross section can still be of the order of $10$ kb. This is surprisingly large  since 
the projectile must irradiate an object with so enormous transverse size \cite{small-large}.  
 
Second, the behaviour of the cross section on the impact energy depends on the value of $R_\perp$: at not very 
large $R_\perp$ the cross section (slightly) decreases with increasing the energy, whereas at very large $R_\perp$ it increases. 

Third, the strength of the relativistic effects 
in the DF mechanism depends both on the impact energy 
and the value of $R_\perp$. 
For instance, at $R_\perp = 52$ \AA \, \,  
the relativistic effects increase the cross section by  
a factor ranging from $1.42$ (at $1$ GeV/u) to $1.82$ (at $10$ GeV/u). With increasing $R_\perp$ the relativistic effects grow and at $R_\perp = 100$ and $200$ \AA \, \, for the same impact energy range this factor varies between $2.69$ and $ 6.24 $,  
and between $ 11.4 $ and $ 44.5 $, respectively. 

At the first sight the above increases might not seem 
dramatic or even especially strong. They, however, are to be compared with the typical strength of relativistic effects in 
collisions with light atoms (or "normal" light molecules). 

For instance, in the single ionization of helium atoms by 
high-energy projectiles the increase of the total cross section 
by a factor of say $1.4$, $6$ and $44$, 
caused by relativistic effects, would be reached 
at impact energies of $\approx 14$ GeV/u, 
$ 8 \times 10^{11}$ TeV/u 
and $ 1.6 \times 10^{105}$ TeV/u, respectively 
\cite{bethe}. 
Moreover, provided ionization is dominated by the independent  interactions of the projectile with each of the "active" target electrons, the total cross section for double ionization of light atoms and molecules is essentially not influenced by relativistic effects at all, no matter how high is the impact energy 
\cite{we-1998}. 

We have found that at $ R_\perp \gg 1$ the total cross section for the production 
of two He$^+$ ions via the DF can quite well be approximated by  
\begin{eqnarray} 
\sigma^{\text{DF}} \approx C \, \frac{ Z_p^4 }{ v^4 \gamma^2 } \, \, \bigg[ K^2_1\left( \frac{ \overline{ \omega } R_\perp }{ \gamma v } \right) + \frac{ 1 }{ \gamma^2 } \, 
K^2_0\left( \frac{ \overline{ \omega } R_\perp }{ \gamma v } \right) \bigg], 
\label{crs-an-DF} 
\end{eqnarray} 
where $ K_0 $ and $ K_1 $ are the modified Bessel function \cite{A-S}, 
$\overline{ \omega } \approx 1.2 $ a.u. is the mean transition frequency for single ionization of a helium atom, $Z_p$ and $v$ are given in a.u. and $ \sigma $ in kb and $ C $ is a parameter weakly dependent on $R_\perp$ \cite{C}. 

\begin{figure}[h!]
\vspace{-0.4cm}
\centering
\includegraphics[width=9.cm]{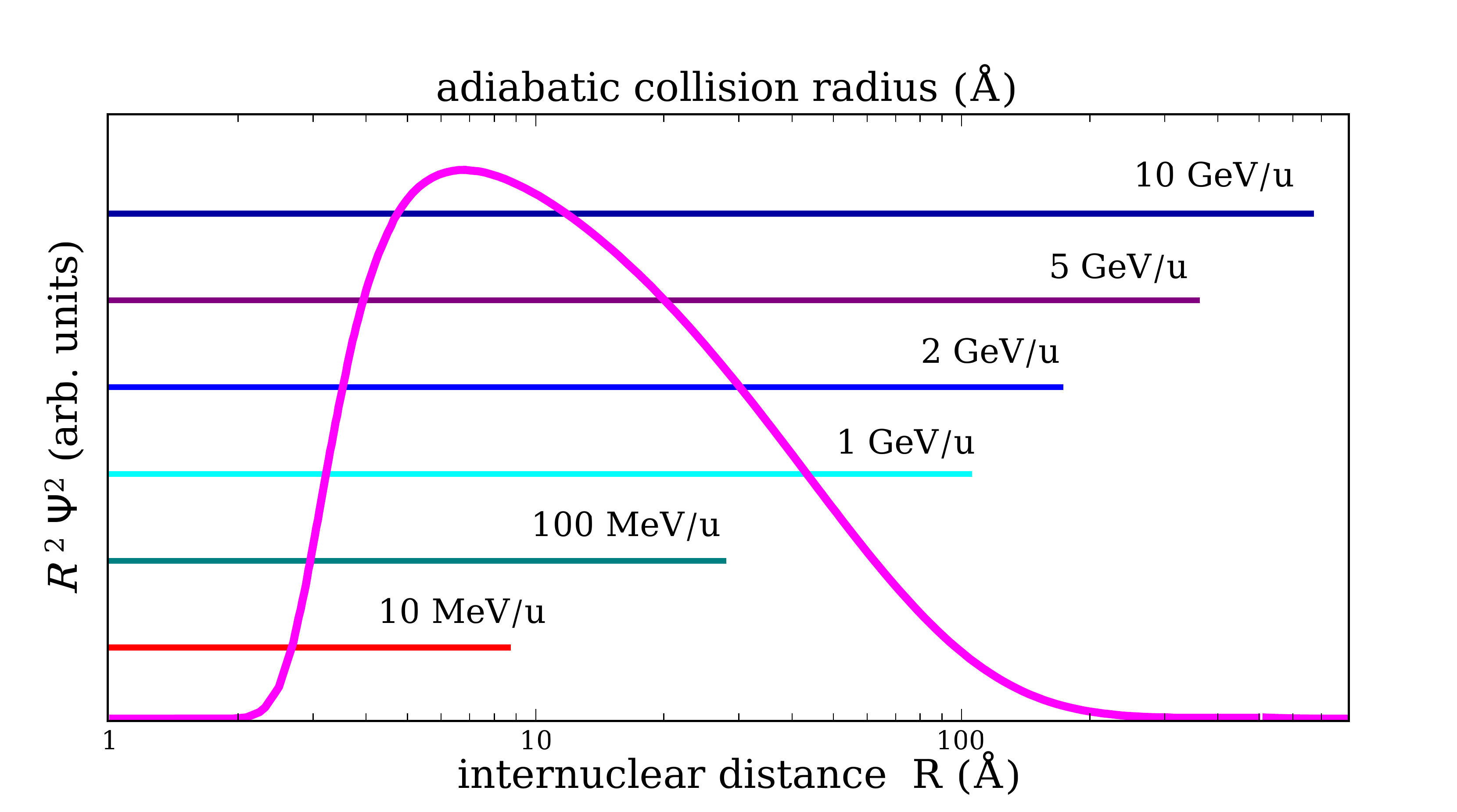} 
\vspace{-0.55cm}
\caption{ The dimer ground state,    
and the adiabatic radius $R_a$ at different impact energies. }
\label{figure3}
\end{figure}

Eq. (\ref{crs-an-DF}) captures all essential features of our numerical results. 
In particular, since at $ x < 1 $ $K_0(x) \sim \ln (1.12/x) $ and $K_1(x) \sim 1/x$ while at 
$ x > 1$ $K_0(x) \sim K_1(x) \sim \sqrt{ \frac{\pi}{2 x }} \exp(-x)$, it is easy to see from Eq. (\ref{crs-an-DF}) that the projectile is able to 
efficiently irradiate both atoms of the dimer 
provided its transverse size $R_\perp$ is smaller 
than the adiabatic collision radius 
$R_a = \frac{ \gamma v }{ \overline{ \omega } } $.  
For an illustration, in figure \ref{figure3} $R_a$ is shown for several impact energies and compared with the extension of the dimer ground state.       

Besides being strongly influenced by relativistic effects, the process of the DF of He$_2$ 
has yet another feature. 
As is known, photo ionization  
by a weak electromagnetic field is a purely quantum process 
while ionization of atoms and 
"normal" molecules by fast charged particles 
can be rather well treated 
by classical physics even in the weak perturbation limit.  
 
However, when $ R_\perp \gg 1$ a.u. 
a classical description of the DF completely fails  
underestimating the cross sections  
by orders of magnitude.    
The reason is that 
very distant inelastic collisions 
are poorely described by a classical treatment
whereas the simultaneous ionization of 
both atoms of the dimer at $ R_\perp \gg 1 $ 
implies that the projectile 
has a very large impact parameter 
with respect to 
at least one of them.    

\begin{figure}[h!]
\centering 
\vspace{0.25cm}
\subfigure{\includegraphics[width=9.cm]{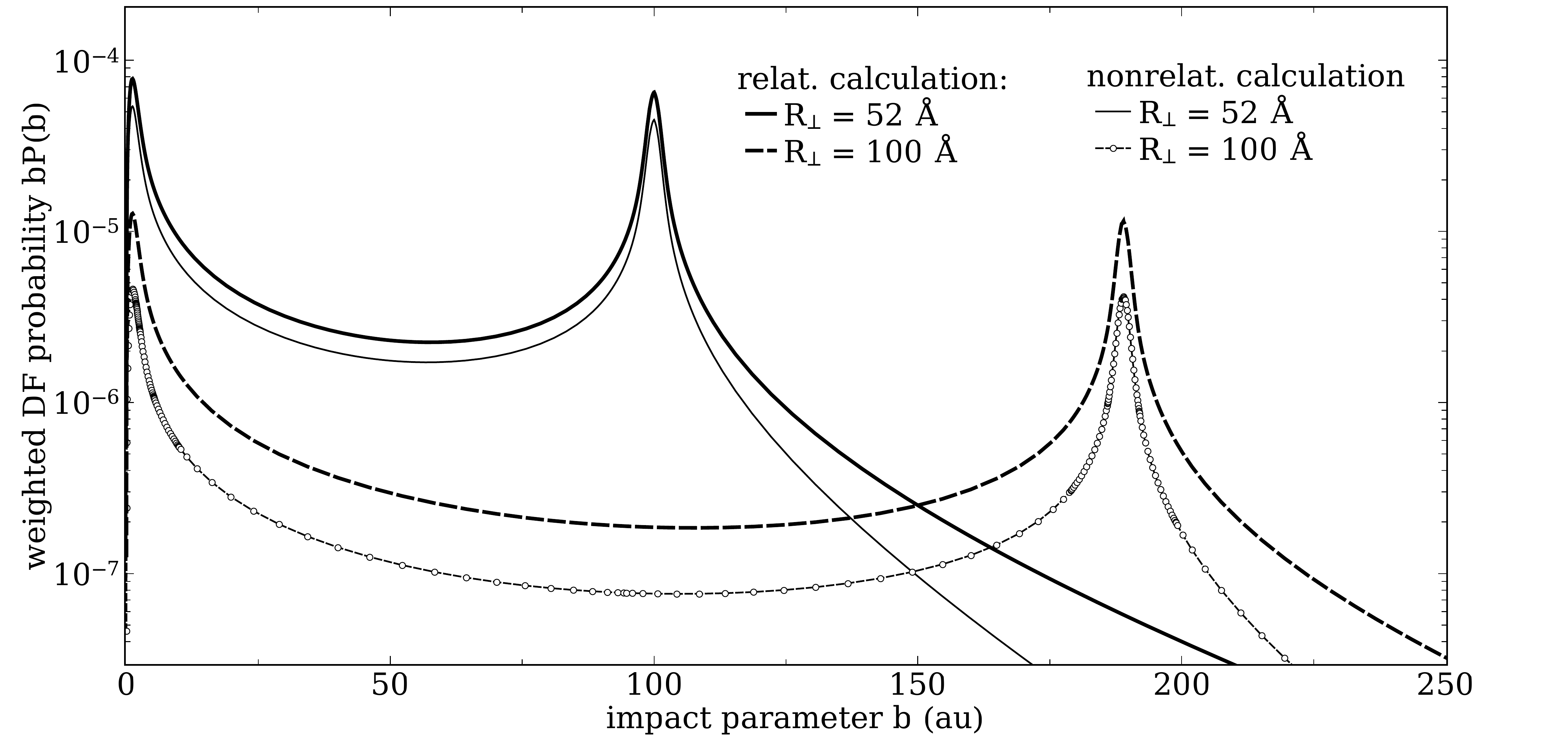}} 
\vspace{-0.55cm}
\caption{ The weighted probability $b P(b)$ 
for the DF given as as function 
of the impact parameter $b$. }
\label{figure4}
\end{figure}

In figure \ref{figure4} we display the weighted probability,  
$b P(b)$, for the DF mechanism as a function of 
the impact parameter $b$ (counted from 
one of the dimer nuclei). It is seen that this probability 
has a pronounced two-peak structure  
showing that the majority of the fragmentation events 
occurs when the projectile passes close 
to either one or the other helium atom.  

6. The SI--e-2e mechanism.  
It will be shown elsewhere that 
in this mechanism the cross section 
for the production of two He$^+$ ions is given by  
\begin{eqnarray}
\sigma^{\text{SI--e-2e}} & = & 
\frac{ 3 \, \sin^2 \vartheta_{\bm R} }{ 4 \pi R^2 } \,   
\int_{I_{He}}^\infty d \varepsilon_k 
\, \frac{ d \sigma^{SI}}{ d \varepsilon_k }  
\, \, \sigma^{\text{e-2e}}(\varepsilon_k) ,    
\label{SI--e-2e} 
\end{eqnarray} 
where $\frac{ d \sigma^{SI}}{ d \varepsilon_k }$ 
is the cross section for single ionization of a helium atom 
by a high-energy projectile differential in the energy $\varepsilon_k$ of the emitted electron, 
$\sigma^{\text{e-2e}}(\varepsilon_k)$ is the cross section for the total single ionization of a helium atom by an electron incident on the atom with an energy $ \varepsilon_k $, $I_{He} \approx 24.6$ eV is the helium ionization potential  
and $\vartheta_{\bm R} = \arccos({\bm R} \cdot {\bm v}/R \, v) $. 

Eq. (\ref{SI--e-2e}) shows that the SI--e-2e mechanism 
is long ranged, depending on $R$ as $ R^{-2}$,   
but becomes inefficient when the angle between 
the dimer orientation  
and the collision velocity is small. 
The dependence of the cross section (\ref{SI--e-2e}) 
on the projectile charge and impact energy is similar 
to that for single ionization of a helium atom:  
in particular, it is proportional to $Z_p^2 $ and weakly 
(logarithmically) influenced by relativistic effects. 

Our analysis shows that in collisions 
with ultrafast ions having very high charges 
the SI--e-2e is much less efficient 
than the DF. However, if $Z_p/v \ll 1$, 
the SI--e-2e becomes dominant 
provided the dimer orientation angle $ \vartheta_{\bm R}$ 
is not too small. 

7. In the DF and SI--e-2e mechanisms 
the Coulomb explosion in the He$^+$-He$^+$ system begins 
when the positions of the dimer nuclei are  
the same as right before the collision. Hence, by measuring 
the $E_{ker}$ spectra produced via these two mechanisms, 
one could directly 
probe the ground state of the dimer   
making its instantaneous "snapshots".  

However, 
in the other two fragmentation mechanisms 
the kinetic energy release  
is not directly related to the dimer size  
at the collision instant. 
Therefore, in order to exclude their interference, 
they have to be "turned off".  
Being characterized by relatively large 
values of $E_{ker}$ ($E_{ker} \gtrsim 1$ eV  
\cite{He2_photon_icd},   
\cite{He2_alpha-particle}), 
they can be "turned off" by simply focusing 
on fragmentation events  
with $E_{ker}$ below $1$ eV. 

In addition to the Coulomb explosion, the He$^+$ ions have kinetic energy from the nuclear motion before the collision:  
it is, however, negligible because the depth of the potential well in He$_2$ is just $1$ meV. Besides,  
the He$^+$ ions also acquire a kinetic energy directly 
in the ionization process.         
In high-energy collisions the momentum transfer $p_{He^+}$ to the He$^+$ ions in an overwhelming majority of ionizing events does not exceed $1$ a.u. \cite{rhci} with the corresponding recoil energy of $2 \times (p^2_{He^+}/2 M_{He}) \lesssim 4 $ meV. 
Therefore, in order that 
the reflection approximation  
$E_{ker} = 1/R$ may still be used,   
one must have $E_{ker} \gg 4$ meV 
(that corresponds to $ R \ll 7 \times 10^3$ a.u.). 

Thus, the DF and SI--e-2e can be used 
for a direct probing of the dimer ground state 
at $ 14 $ \AA\,  $ < R \lesssim 250 $ \AA  \, 
(corresponding to 
$ 60 $ meV $ \lesssim E_{ker} < 1$ eV), i.e. 
of its halo region where the dimer spends most of the time.   
\begin{figure}[h!]
\vspace{0.0cm}
\centering
\subfigure{\includegraphics[width=9.5cm]{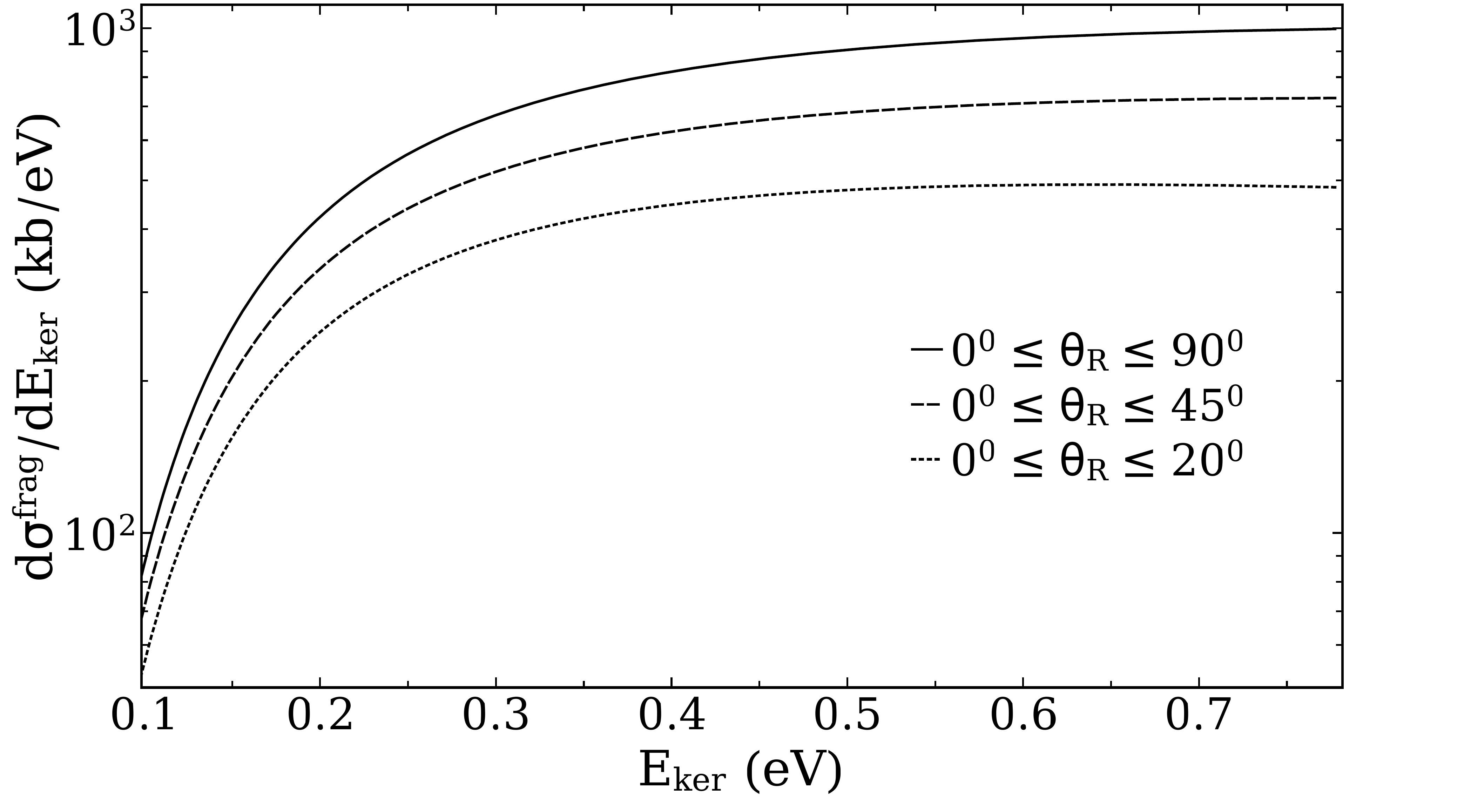}} 
\vspace{-0.55cm}
\caption{ Kinetic energy release spectra. }
\label{figure5}
\end{figure}
Using the cross section for the production of two helium ions and the wave function of the dimer ground state we can calculate the dimer fragmentation cross section $\frac{ d \sigma^{\text{frag}} }{ d E_{ker} }$ differential in the kinetic energy release. 
In figure \ref{figure5} we present it for collisions with $5$ GeV/u U$^{92+}$ projectiles,   
for different ranges of the dimer orientation. 
Because of the extremely long interaction range inherent 
to the ultrafast projectile, the spectrum intensities remain significant even at 
energies corresponding to very large internuclear distances $R$. 

At $R \gg 1$ the probability density of the dimer is 
$\vert \Psi \vert^2 \sim \exp(- 2 \kappa_0 \, R)/R^2$,  where $\kappa_0 = \sqrt{ M_{He} \, I_b/\hbar^2 }$ 
with $ M_{He}$ being the mass of the helium atom.    
A small variation $\Delta I_b$ of the dimer binding energy $I_b$ ($\Delta I_b/I_b \ll 1 $) changes 
$\vert \Psi \vert^2 $ 
by the factor $\exp(- \kappa_0 R \, (\Delta I_b/I_b)$). 
This in turn affects the shape of the kinetic energy release spectrum. Since this spectrum effectively spans  
a very broad range of $R$, it becomes sensitive even to  
a very small change in the dimer binding energy. 
 
In particular, the reported values for the binding energy vary 
between $44.8$ neV \cite{44.8} and $161.7$ neV \cite{161.7}, with $139.2$ neV \cite{139.2} 
(used in our calculations) 
and $151.9$ neV \cite{dimer-binding-exp} being 
regarded as most precise having the relative difference 
of just $ 9 \%$. 
However, in the ranges 
$ 0.1 $ eV $ \leq E_{ker} \leq 0.8$ eV, 
$ 0.075 $ eV $ \leq E_{ker} \leq 0.8$ eV 
and $ 0.06 $ eV $ \leq E_{ker} \leq 0.8$ eV 
(the latter two are not shown in fig. \ref{figure5})  
this $9 \%$ are already converted into, respectively,  
$14 \%$, $ 20 \% $ and $ 26 \%$ difference in the shape of the energy spectrum. This suggests that collisions with ultrafast projectiles can be used for an accurate determination of 
the He$_2$ binding energy.        

The cross section for the DF, integrated over the kinetic energy release between $0.1$ and $1$ eV by U$^{92+}$ projectile having impact energy 
$\sim 1$-$5$ GeV/u is of the order of 
$10^{-19} - 10^{-18}$ cm$^2$.  This value 
could be compared with the total cross section
for single ionization of helium atom, which is 
$\sim 10^{-15}$ cm$^2$, the cross section for single electron capture from helium atom, which is $\sim 10^{-23}$ cm$^2$  
and the cross section 
for the reaction $Z_p$ + He$_2$ 
$ \to $ $Z_p$ + He(1s$^2$) + He(1s$^2$), which (according to rough estimates) 
is $\sim 10^{-16}$ cm$^2$.
  
8. In conlusion, we have studied    
the fragmentation of the helium dimer into 
singly charged helium ions by 
relativistic highly-charged projectiles. 
It was found that the breakup events 
with kinetic energy release $< 1$ eV  
are in essense solely  
caused by the direct fragmentation mechanism 
in which the projectile "simultaneously"  
ionized both dimer's atoms. 
It was shown that this mechanism 
is exceptionally strongly influenced by relativistic effects 
and that a classical description of the collision dynamics in this case completely fails. 

It was also demonstrated that ultrafast projectiles, 
due to their extremely long interaction range,   
can directly and accurately probe 
the structure of the dimer ground state 
in the halo region $\sim (14 - 250) $ \AA \, 
where the dimer spends four-fifths of the time. 
Moreover, a rather high sensitivity of the $E_{ker}$-spectrum 
to the value of the dimer binding energy suggests that such projectiles can be used for its more precise determination. 

Collisions with ultrafast projectiles can also be applied to explore the ground states of $^6$LiHe and $^7$LiHe dimers, 
which are other humungous diatomic molecules having the average size of about $ 49 $ and $ 28 $ \AA, respectively, \cite{LiHe}.  

Finally we note that  
experimental facilities, 
where very large dimers can 
be explored by collisions 
with ultrafast highly charged projectiles, 
are available.

Acknowledgements. 
The authors gratefully acknowledge the support from the China Scholarship Council and CAS President’s Fellowship Initiative.


\begin{thebibliography}{33}
\bibitem{el-cpt} 
For instance, in collisions of 
$1$ GeV/u U$^{92+}$ with a helium atom 
the cross section for radiative electron capture 
is roughly by $8$ orders of magnitude smaller 
than the cross section for helium single ionization 
(the cross section for nonradiative capture 
is even much smaller).
\bibitem{Sulf} At much lower impact energies this mechanism was considered in \cite{He2_S14+} for collisions with helium dimers having the internuclear distance in the range $[5, 10]$ a.u.
\bibitem{RA} Note, in particular, 
that the application of the reflection approximation 
enabled the authors of 
experiment \cite{dimer-binding-exp} to 
explore the dimer ground state and obtain 
an accurate value for its binding energy.
\bibitem{f1} For the fragmentation of He$_2$ 
by a photon a similar mechanism was experimentally explored 
in \cite{He2_photon}.
\bibitem{Z-v} 
Due to very high impact energies,  
even for $Z_p = 92$
the parameter $\eta = Z_p/v$ varies between 
$0.77$ and $0.67$ showing that the effective field of the projectile in the collision is rather weak than strong.
\bibitem{small-large} Note that the cross section 
for the He$_2$ fragmentation at $R = 500 $ a.u. by relativistic U$^{92+}$ projectiles are comparable in magnitude to those for the $K$-shell ionization of the heaviest atoms by the same U$^{92+}$ projectiles.
\end{thebibliography}
\end{document}